# Unsupervised CNN-Based DIC for 2D Displacement Measurement

Yixiao Wang [a], Canlin Zhou [*a], Si ShuChun [a], Hui Li [a]

a School of Physics, Shandong University, Jinan 250100, China

* canlinzhou@sdu.edu.cn; phone :+86 13256153609

**Abstract:** Digital image-correlation（DIC） method is a non-contact deformation measurement technique. Despite years of development, it is still difficult to solve the contradiction between calculation efficiency and seed point quantity.With the development of deep learning, the DIC algorithm based on deep learning provides a new solution for the problem of insufficient calculation efficiency in DIC.All supervised learning DIC methods requires a large set of high-quality training set. However, obtaining such a dataset can be challenging and time-consuming in generating ground truth. To fix the problem,we propose an unsupervised CNN-Based DIC for 2D Displacement Measurement.The speckle image warp model is created to transform the target speckle image to the corresponding predicted reference speckle image by predicted 2D displacement map, the predicted reference speckle image is compared with the original reference speckle image to realize the unsupervised training of the CNN.The network's parameters are optimized using a composite loss function that incorporates both the Mean Squared Error (MSE) and Pearson correlation coefficient.Our proposed method has a significant advantage of eliminating the need for extensive training data annotations. We conducted several experiments to demonstrate the validity and robustness of the proposed method. The experimental results demonstrate that our method can achieve can achieve accuracy comparable to previous supervised methods. The PyTorch code will be available at the following URL: https://github.com/fead1.

**Keywords:** digital image-correlation, digital speckle image, two-dimensional displacement measurement , convolutional neural network,unsupervised learning,training data, ground truth,Pearson correlation coefficient,Mean Absolute Error, loss function

1.Introduction

DIC(digital image-correlation) method is a non-contact deformation measurement technique. Due to its advantages of full-field measurement, simple optical path and high accuracy, it has become an important means of displacement measurement. The DIC algorithm framework can be divided into local subregion DIC algorithm and global DIC algorithm [1]. Despite years of development, it is still difficult to solve the contradiction between calculation efficiency and seed point quantity. With the development of deep learning, the DIC algorithm framework based on deep learning provides a new solution for the problem of insufficient calculation efficiency in DIC. In 2020, Min et.al.[2] used a three-dimensional convolutional neural network to achieve end-to-end deformation field output, avoiding interpolation and optimization processes in traditional DIC algorithms and directly outputting full-pixel displacement fields but with limited measurement accuracy. Boukhtache et.al.[3-4] trained neural networks for speckle image using optical flow computation domain to achieve sub-pixel matching accuracy; however this method did not consider whole pixel matching and required traditional methods to provide initial value guesses. To measure large deformed objects, Yang et.al. proposed using two encoding-decoding networks respectively calculate strain and displacement of large deformed objects by updating regions-of-interest (ROIs) for measuring large deformed objects, Zhao Jian et.al.[5] proposed using image classification networks for sub-region matching but matching accuracy was limited by number of classification types with limited generalization ability. Subsequently they improved their network as regression network only used for initial guess in DIC[6]. Huang et.al. did similar work[7], they still adopted local subregion DIC algorithm framework so calculation efficiency and tracking seed point quantity were still limited. Zhang et.al.[8] used U-Net network for deformation field measurements where calculating pixels' deformation field from a pair of speckle image with size of 512x512 only took 4.3 milliseconds but this method was not suitable for large deformations greater than eight pixels. Zhao et.al. [9] proposed a new Hermite dataset that utilizes high-order Hermite elements to account for more complex deformation modes. The residual block is used to encode the path for extracting additional features, resulting in DIC measurements with an accuracy of up to 0.013 pixels. However, this method only considers micro-deformations where the deformation amount is less than one pixel.

All of the DIC methods based on deep learning mentioned above use a supervised learning approach. To train the network using this method, a large set of high-quality training data is required. However, creating such a dataset can be time-consuming and challenging due to algorithmic defects and difficulties in generating ground truth. Therefore, obtaining an efficient training model with high-quality data remains an unsolved problem in this field. Model-driven learning methods offer an alternative solution by solving inverse problems without labels through neural networks' interaction with physical models. These methods have been successfully applied to autofocus [10], 3D reconstruction[11],image fusion[12],hyperspectral recovery[13] and computer-generated holography [14].

Inspired by these successful applications of unsupervised learning. We propose the establishment of an unsupervised framework for training the Digital image-correlation Neural Network for two-dimensional displacement measurement. To our knowledge, this is the first time that unsupervised learning has been applied to Digital Image-Correlation for displacement measurement. To successfully implement an unsupervised deep learning network, it is crucial to establish a mathematical relationship between the input speckle image data and the predicted two-dimensional displacement data. This approach involves connecting digital speckle images before and after displacement while taking into account physical constraints. After predicting the 2D displacement map, the speckle image can be displaced accordingly. The inversion model will then re-project this displaced speckle image onto its original state before displacement. This allows for a comparison between the newly created speckle image and the input speckle image before displacement, which is necessary to establish loss for training the network. The network parameters are optimized by a composite loss function that combines the Mean Squared Error (MSE) and Pearson correlation coefficient to calculate the loss. Our proposed network eliminates the need for extensive data annotations and can be trained without ground truth, making it suitable for unsupervised learning. Our experimental results demonstrate that our method achieves competitive accuracy compared to previous supervised methods while also saving time on dataset labeling.

**2. Method**

In this section, we propose a DIC method based on unsupervised convolutional neural network for displacement field measurement. After using a small amount of ground truth to conduct supervised pre-training on the network, this method can only use reference speckle and target speckle patterns when there is no ground truth data.Using unsupervised learning mode for network training.In Section 2.1, we introduce the realization method of using displacement field prediction in unsupervised convolutional neural network .Section 2.2 shows our proposed network structure and the loss function.

## 2.1 Basic Theory

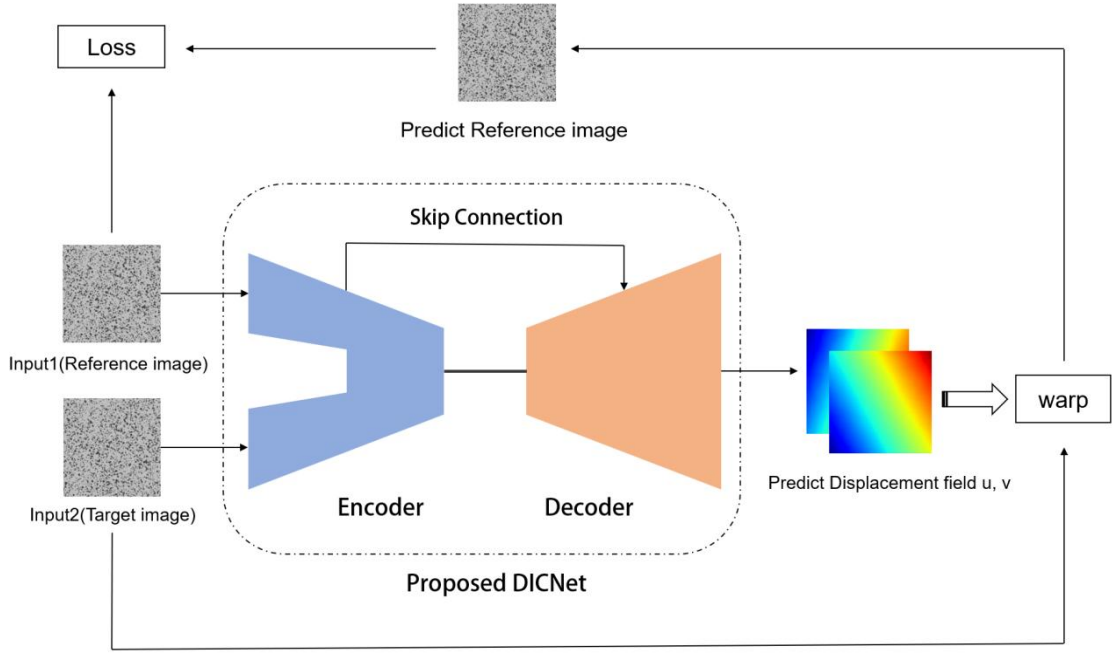

Fig.1 The unsupervised network architecture

As shown in Figure 1, the collected reference image before deformation and the target image after deformation are separately input into our proposed DICNet through two branches. At the Encoder stage of the network, deep information from both images is extracted separately, and then associated and fused. In the Decoder stage, upsampling and reconstruction are performed to produce a displacement field with the same size as the input images. The displacement field predicted by the network contains two-dimensional displacement vectors (u, v), representing displacement in the x and y directions, respectively.Based on the working principle of digital speckle correlation, the target image after deformation is obtained by applying the displacement to the reference image. Therefore, by inverting the predicted displacement field and applying it to the target image, the predicted reference image can be obtained. To complete the unsupervised training of the network, the predicted reference image is connected with the real reference image, and the loss function of the network is constructed.In practical deformation and displacement, the displacement may be sub-pixel rather than integer. When inverting the target image using the displacement, the gray data values of integer pixel positions in the predicted reference image can be obtained by a bilinear cubic interpolation method.

By using the actual physical constraint relationship between the reference speckle pattern and target speckle pattern before and after deformation in speckle correlation, the network's predicted displacement field can be applied to the target speckle image, and the deformation process of the target speckle image can be inverted to obtain the predicted reference speckle image. The predicted reference speckle image and the actual reference speckle image are connected to construct the network's loss function, enabling supervised learning of the network.

It is important to note that in the early stages of network training, the network parameters are typically initialized randomly. This may cause unsupervised training of the network to converge in the wrong direction, as certain pixel locations in the target image may have similar gray values to multiple locations in the reference image. To address this issue, we conducted 30 epochs of supervised pre-training using a dataset containing a small number of real displacement fields before unsupervised network training.

After using a small set of specific 2D displacement map ground truth data for network pre-training to constrain the learning direction of the network, the framework is able to learn 2D displacement in an unsupervised manner that doesn't require massive amounts of supervised ground truth, and thereby facilitates much faster data labeling compared to purely supervised frameworks. When the loss function stops decreasing, unsupervised training is complete, and the trained network can be used for 2D displacement reconstruction.

## 2.2 Network structure and loss function

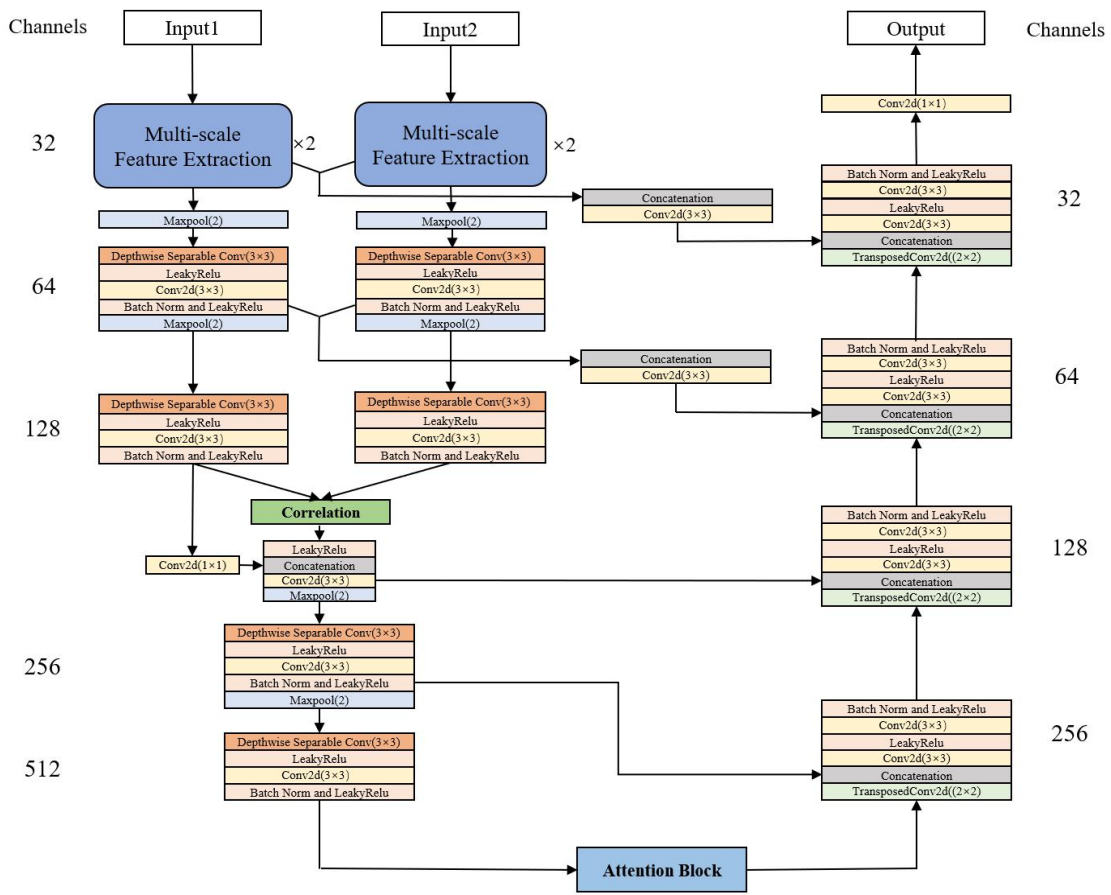

Fig.2 The detailed architecture of the proposed DICNet

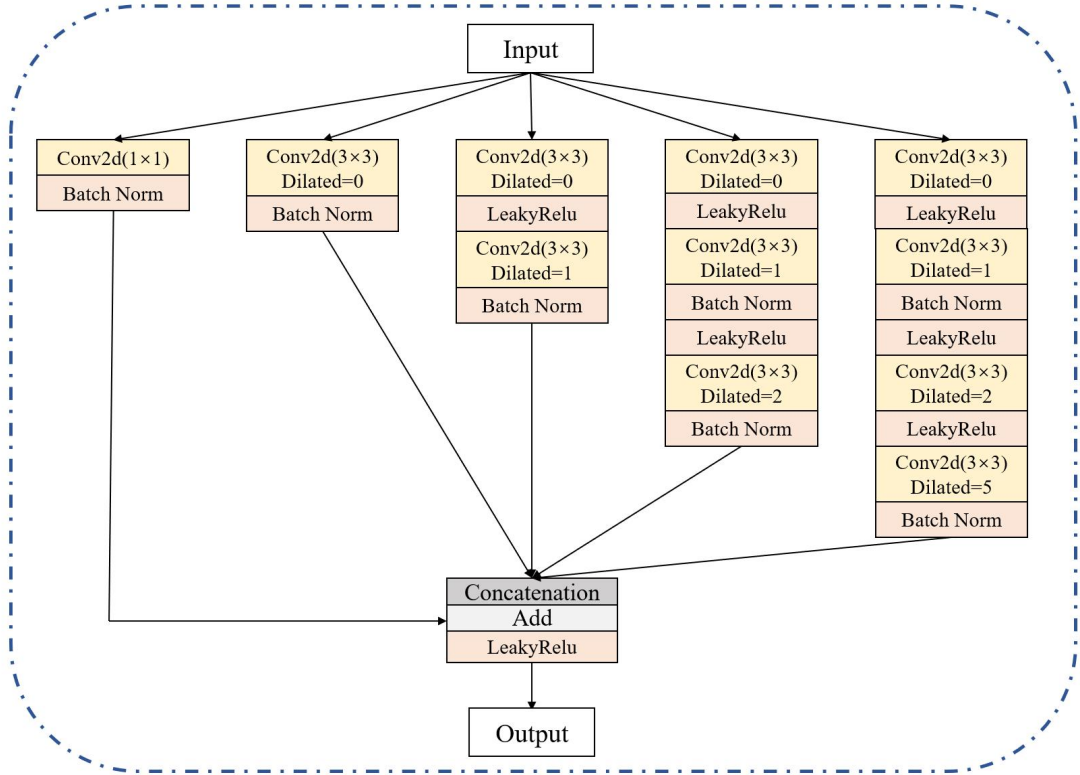

Fig.3 Multi-scale Feature Extraction Block

As shown in Figure 2, our proposed DICNet is an end-to-end displacement prediction network that was inspired by Unet[15] and FlowNetCorr[15]. The network structure consists of an encoder-decoder architecture similar to UNet, and a dual-channel input structure with a correlation layer similar to FlowNetCorr. The encoder-decoder structure has excellent information extraction and reconstruction capabilities, effectively extracting and mapping the deformation information contained in the images to a displacement field. The dual-channel input structure extracts deep features from both images and then combines and matches them to improve the feature extraction capability in the encoding stage. The correlation layer uses the correlation between two feature maps to allow the network to learn and represent the displacement information contained in the two images before and after deformation more efficiently and accurately. Additionally, an attention mechanism was added to the encoder stage to further enhance the network's feature correlation and fusion capabilities.

In Figure 2, input1 and input2 are respectively the reference and target images from Figure 1, with input formats of $B \times 1 \times H \times W$, where the first dimension B represents the batch size of the input images, the second dimension represents the number of channels (1 for gray-scale images), and the third and fourth dimensions represent the height and width of the input images, respectively.

Input1 and Input2 are first processed by separate branches of the network to extract deep-level feature information. Due to the diverse range of displacement deformation patterns present in the input dataset, with different reference and target images having varying displacement patterns and magnitudes, we designed a multi-scale feature extraction module and placed it in the first layer of the network to complete the extraction of features across different scales and depths of the input images.

The multi-scale feature extraction module, shown in Figure 3, divides the input feature map into five branches, with the right four branches using different numbers of dilated convolutions with varying dilation rates. The multiple layers of dilated convolutions enable the extraction of features at different depths of the input image, while the various dilation rates and convolution depths enable the module to adapt to displacement information of different sizes present in the input image. The feature maps from the four branches are merged through concatenation, and the leftmost branch serves as a skip connection, adding the features of the input image to the merged feature maps to ensure smoother gradient back-propagation.

In Figure 2, after passing through the two multi-scale feature extraction modules, the feature maps proceed to the subsequent feature extraction tasks using a mix of depthwise separable convolutions with 3×3 convolution kernels and regular convolutions. At each convolution layer, we apply LeakyRelu or BatchNormal+LeakyRelu operations to the network for regularization and activation. The use of depthwise separable convolutions compared to regular convolutions greatly reduces the number of network parameters, resulting in a more lightweight network architecture.

After extracting features from the reference and target images, the dual-channel network performs correlation operations at a higher level. Specifically, feature vectors are extracted from patches of the two feature maps and compared against each other. Given the two feature maps, we adopt the "correlation layer" method proposed in FlowNetcorr[15], enabling the network to identify the correspondences between the feature maps.

The correlation operation has been marked in Figure 2. For the two feature maps, f1 and f2, the correlation layer establishes correlations between every patch in f1 and every patch in f2. For example, given two patches from f1 and f2 with centers at x1 and x2, respectively, the correlation operation is defined as follows:

$$c(\mathbf{x}_1, \mathbf{x}_2) = \sum_{\mathbf{o} \in [-k,k] \times [-k,k]} \langle \mathbf{f}_1(\mathbf{x}_1 + \mathbf{o}), \mathbf{f}_2(\mathbf{x}_2 + \mathbf{o}) \rangle \qquad (1)$$

In the above equation, k is half of the window size corresponding to each patch.
By combining equation (1) with Figure 4, it becomes apparent that the correlation operation essentially calculates the inter-correlation between two patches in f1 and f2 by convolving with patches of the same size K2 (K=2k+1) in f2 using the patches in f1 as a convolution kernel.

Equation (1) shows that computing the correlations between all patches in f1 and f2 requires a calculation complexity of $W^2 \times H^2$, where W and H are the widths and heights of the feature maps f1 and f2, respectively. Such computational complexity would be prohibitive for training the network and for model light-weighting. Therefore, we introduce the maximum displacement limit d to address this issue. As shown in Figure 4, we compute the correlations by limiting x2 to the neighborhood of x1 with a range of D (D=2d+1), instead of across the entire image. This reduces the computational complexity of the correlation operation.

After calculating $c(\mathbf{x}_1, \mathbf{x}_2)$ within the limited D range around a specific x1, we move x1 through f1 row by row and column by column, repeating the same process to obtain the correlation feature maps between f1 and f2. Correlation is calculated pixel by pixel, presuming that the shapes of the feature maps F1 and F2 of the two branches entering the correlation operation in Figure 2 are C0 × H0 × W0. The restriction range is set to D. Thus, a correlation matrix of size $D^2$ will be produced. After two feature maps F1 and F2 are correlated, a four-dimensional tensor of the shape H0 × W0 × D × D is produced. The correlation feature map can then be obtained by reshaping it into the shape of $D^2$ × H0 × W0. It should be noted that d can be estimated and adjusted based on the actual displacement. For instance, if the data set contains speckle images with larger displacement, d should be increased, while for smaller displacement, d can be reduced. Reasonable setting of the d value can not only extract displacement features through correlation operation, but also reduce computational complexity.

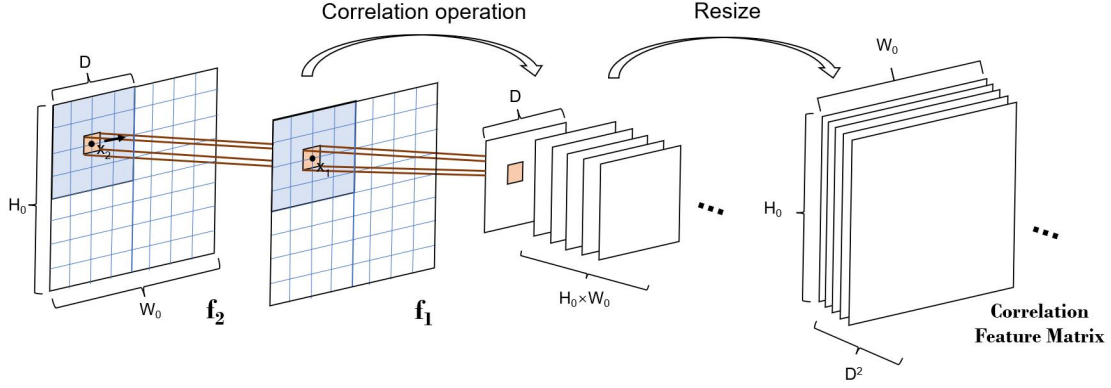

Fig.4 correlation layer

As shown in Figure 2, we introduce an attention module at the end of the encoding stage to further enhance the network's ability to extract effective information for representation. Our network structure employs a hybrid attention module that includes both channel attention and spatial attention, as shown in Figure 5. The channel attention module is implemented according to equation (2). Given an input feature map F (in the form of C×H×W), it first undergoes global maximum and average pooling operations along the spatial dimension to obtain two channel descriptors of size C×1×1, which are then fed into a two-layer shared neural network. The two features obtained from the shared network are then added together and passed through a Sigmoid activation function to obtain weight coefficients Mc. Multiplying Mc with the input feature map reshapes its channels, which attenuates the "unimportant" channels and strengthens the "important" channels.

The spatial attention module is implemented according to equation (3), which is similar to the channel attention module. Given an input feature map F of size C×H×W, it undergoes maximum and average pooling operations along the channel dimension to obtain two spatial descriptors of size 1×H×W, which are then concatenated along the channel dimension. The resulting feature map is convolved with a 7×7 kernel and passed through a Sigmoid activation function to obtain weight coefficients Ms. Multiplying Ms with the input feature map reshapes its spatial positions, completing the network's attention learning for spatial positions.

$$M_c(F) = \sigma(MLP(\text{AvgPool}(F)) + MLP(\text{MaxPool}(F))) \qquad (2)$$

$$M_s(F) = \sigma\left(f^{7*7}([\text{AvgPool}(F), \text{MaxPool}(F)])\right) = \sigma\left(f^{7*7}\left([F_{\text{avg}}^S; F_{\text{max}}^S]\right)\right) \qquad (3)$$

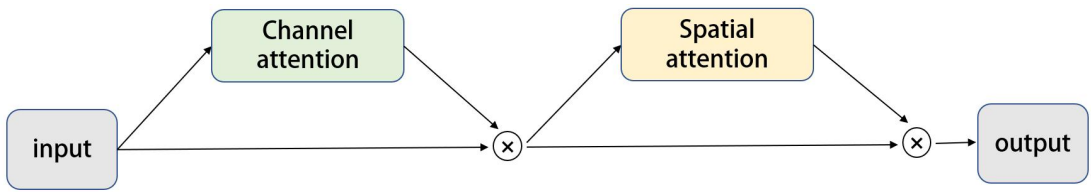

Fig.5 Hybrid attention Block

The addition of the hybrid attention module further enhances the network's representation ability. The channel attention module strengthens the network's focus on channels that contain more displacement field information, allowing the network to learn the displacement field information more quickly and effectively. The spatial attention module changes the network's focus on different locations in the feature map, improving the network's prediction accuracy for complex displacements.The network predicts the corresponding displacement field based on a pair of speckle images before and after deformation. In unsupervised training, how to construct a loss function based on the predicted displacement data and the actual physical constraint relationship between the reference and target speckle image is a key issue in unsupervised learning.We apply the predicted displacement field to the target image to obtain the predicted reference image. The process is shown in Figure 6. First, we create an initial grid matrix grid0 that is the same size as the target image, where x and y represent the horizontal and vertical directions, respectively, as shown in Figure 6(a). The values at the same index of the two grid matrices form a coordinate distribution (x0, y0). In the initial grid matrix x, the value x0 at the index (x, y) is the same as x. Similarly, in the initial grid matrix y, the value y0 at the index (x, y) is the same as y.By adding the initial grid matrix grid0 and the predicted displacement field of the network, we can obtain the required sampling matrix grid.To ensure that gradients can be smoothly backpropagated during network training, we need to perform sampling and interpolation operations on tensor data. Therefore, we use the built-in function 'grid_sample' [17]in PyTorch instead of the 'griddata' function in the Scipy library (which cannot handle tensor data) to perform the sampling work of grid in the target image.

The procedure of 'grid_sample' function is roughly shown in Figure 6(b). Assuming that the values of the two sampling grid matrices x and y at position (x, y) are x1 and y1, respectively, the function will obtain the grayscale value at position (x1, y1) in the target image and place it at position (x, y) in the predicted reference image. By processing the entire grid in a similar manner, the predicted reference image corresponding to the target image can be obtained. In this process, if the sampling position of the grayscale value in the target image is a non-integer pixel, bicubic interpolation is used to predict the grayscale value at the sampling point. It should be noted that before sampling, the values of the sampling grid matrix grid need to be normalized to the range of [-1, 1], and then grid and target image are inputted to the 'grid_sample' function to obtain the predicted reference image. In addition, there is an important parameter 'align_corners' in 'grid_sample'. If set to 'True', the extrema ('-1' and '1') are considered as referring to the center points of the input's corner pixels. If set to 'False', they are instead considered as referring to the corner points of the input's corner pixels. Through experiments, we found that setting it to 'True' will result in more accurate sampling results.

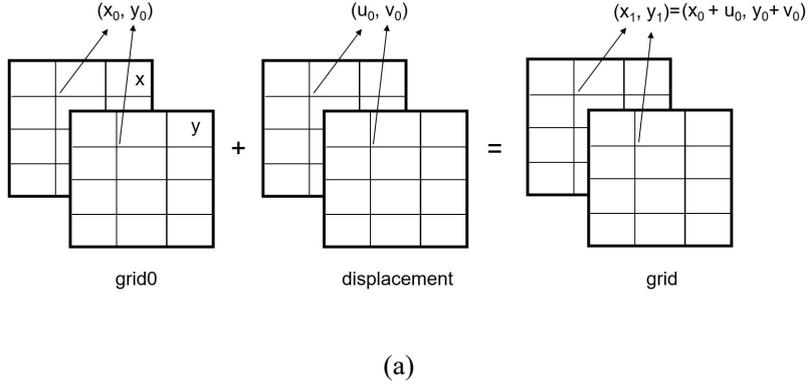

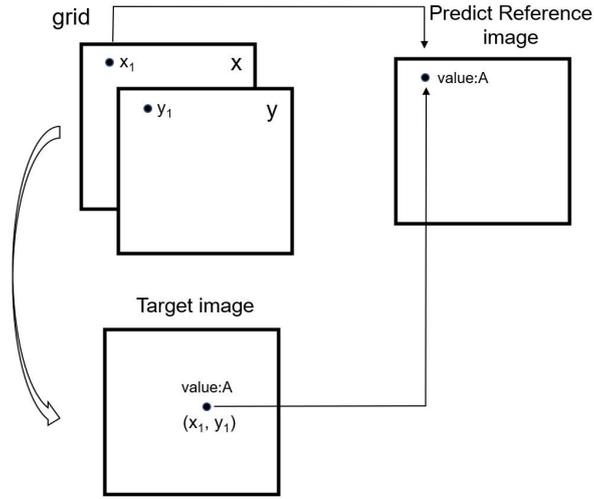

Figure 6.The process to obtain the predicted reference image.

(a) Generation of sampling grid. (b) Warping process of the target image.

By using the above methods to obtain the predicted reference image, we can design a loss function to train the network. In the design of the loss function, we employ the classic MSELoss function in the supervised learning process to conduct network pre-training. Due to the easy differentiability of the MSELoss and its ability to effectively penalize large prediction errors, the network can converge more efficiently during the pre-training stage.

In the unsupervised learning, we combine the MSELoss with the Pearson correlation coefficient loss (losscorr) to form a hybrid loss function, which is given by the following form.

$$Loss_{total} = loss_{MSE} + loss_{corr} \quad (4)$$

The lossMSE is the mean squared error, which is expressed as follows:

$$loss_{MSE} = \frac{1}{n}\sum_{i=1}^{n}(y_i - y_i^p)^2 \quad (5)$$

where n is the total number of image pixels, $y_i$ represents the grayscale value of the reference image, and $y_i^p$ is the grayscale value of the predicted reference image.

During the unsupervised learning stage, the loss function connects the reference image and the predicted reference image. Although lossMSE can make the network converge easily, if the dataset contains target images that are overall brighter or darker than the reference images due to experimental factors, the use of a single lossMSE can lead to incorrect convergence. For this reason, we added losscorr, which is expressed as follows:

$$\text{loss}_{\text{corr}} = 1 - \rho_{y,y^p} \tag{6}$$

$\rho_{y,y^p}$ is the Pearson correlation coefficient between the reference image and the predicted reference image, and is expressed as follows:

$$\rho_{y,y^p} = \frac{\text{Cov}(y, y^p)}{\sigma_y \cdot \sigma_{y^p}} = \frac{\sum_{i=1}^{n}(y_i - mean(y))(y_i^p - mean(y^p))}{\sqrt{\sum_{i=1}^{n}(y_i - mean(y))^2} \cdot \sqrt{\sum_{i=1}^{n}(y_i^p - mean(y^p))^2}} \tag{7}$$

where $y_i$, $y_i^p$ represent the grayscale values of the reference image and the predicted reference image while $mean(y)$ and $mean(y^p)$ represent the mean grayscale values of the reference image and the predicted reference image, respectively.

The addition of losscorr effectively reduces the network's sensitivity to outliers, and enhances the network's generalization ability and robustness. In our experiments, we also found that using Losstotal as the loss function, instead of using a single lossMSE, led to better convergence results for the network.

### 3. Experiments

In this section, we conducted experiments to validate the effectiveness of our proposed DICNet and unsupervised neural network for displacement field measurement. We used an open-source dataset [] that contained 4000 samples, with 2400, 1200, and 400 samples in the training, validation, and testing sets, respectively. The sample size was 480 pixels × 480 pixels, and the dataset included a large amount of data with significant displacement. The maximum displacement in the dataset was up to 16 pixels.

The experiment uses a desktop computer with an Intel Core i7-9700k processor, a 32-GB RAM, and a Nvidia GeForce GTX 2080Ti. The code for training is written in Pytorch and utilizes the Adaptive Moment Estimation (Adam) optimizer.

In this experiment, we compared our proposed method with Zhao et al.'s method. The comparison of the two method in terms of Number of parameters, Memory size and inference speed is shown in Table 1.

Table 1: Comparison of parameters between the two methods.

| Model | Number of parameters(M) | Memory size (MB) | Speed(s) |
|---|---|---|---|
| Zhao method | 37.886 | 144.52 | 0.0678 |
| Our method | 7.274 | 27.75 | 0.0390 |

From Table 1, it can be seen that compared to Zhao et al.'s method, our method has the advantages of having fewer network parameters, requiring less memory size for the model, and faster inference speed. For the above two methods, we adopted three training modes, namely A, B, and C, to compare their performance on the testing set. The three training modes are as follows:

A Mode: Using the method proposed by Zhao et al., the entire training set was used for 200 epochs of supervised training.

B Mode: Using our proposed DICNet, the entire training set was used for 200 epochs of supervised training.

C Mode: Using our proposed DICNet, the first 1/24 samples (100 sample ) of the entire training set were used for 30 epochs of supervised training, and then the entire training set was used for 170 epochs of unsupervised training.

After training, we tested the network models trained by three training modes on the same testing set. Due to the large displacement in the samples in this dataset, there may be partial displacement information missing around the sample images. Therefore, when analyzing the errors and displaying the results, we cropped a 440 pixel × 440 pixel square region of the displacement fields outputted by different models from the center for comparison.

Here, we use the MAE and RMSE metrics to calculate the errors between the predicted displacement fields and the ground-truth. They are formulated as follows:

$$\text{MAE} = \frac{1}{n}\sum_{i=1}^{n} |y_i - y_i^p| \tag{8}$$

$$\text{RMSE} = \sqrt{\frac{1}{n}\sum_{i=1}^{n} (y_i - y_i^p)^2} \tag{9}$$

Where $y_i$ is ground-truth, $y_i^p$ is predicted result by network, n represents the number of valid data points in the displacement field.

The networks trained by A Mode, B Mode, and C Mode were tested on the testing set to compare the errors. The results are shown in Table 2.

Table 2:Comparison of errors of different modes .

| Mode | MAE(pixels) | RMSE(pixels) |
| --- | --- | --- |
| A | 0.160 | 0.199 |
| B | 0.0665 | 0.0853 |
| C | 0.0637 | 0.0828 |

From Table 2, comparing Mode A and B, we can see that under the same training conditions , our proposed DICNet has reduced the MAE and RMSE errors of the predicted displacement fields on the testing set by 58.44% and 57.14%, respectively, compared to Zhao et al's method . This indicates that our proposed DICNet can more accurately handle test samples with large displacements and precisely predict the displacements. In Table 2, Mode B and C both use the same network architecture(proposed DICNet) , but are trained with supervised and unsupervised learning, respectively. The MAE and RMSE errors between the predicted displacement and ground-truth by the unsupervised learning are only 0.0637 pixels and 0.828 pixels, respectively. The prediction accuracy is even slightly higher than that of supervised training mode.

To further demonstrate the displacement field prediction ability of the networks trained by the three modes, we selected four sample images of different types from the testing set and visualized the predicted displacement fields, as shown in Figure 7. Combined with the error comparison data in Table 2 and the errors between the predicted displacement and ground-truth , it can be seen that our proposed DICNet has higher prediction accuracy and stability compared to the method proposed by Zhao et al. Additionally, comparing B Mode and C Mode, we can see that our proposed unsupervised neural network did not use the ground-truth labels during training after pre-training using a dataset containing a small number of real displacement fields, but the predicted  displacement  results on the testing set are comparable to those of supervised training.

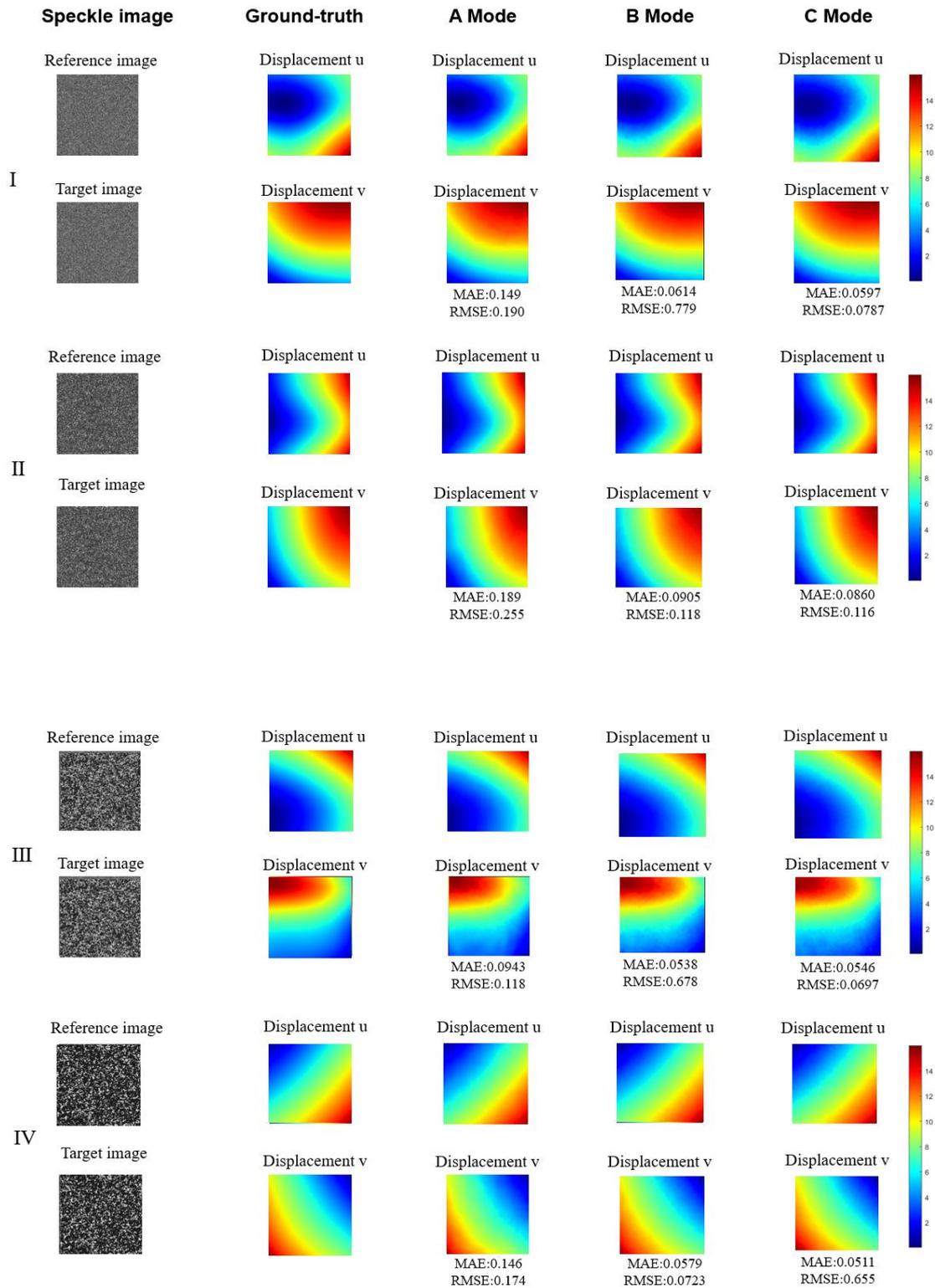

Fig.7, The predicted displacement fields from four different samples processed by different training Mode

Ⅰ row   for sample 1; Ⅱ row   for sample 2; Ⅲ row   for sample 3; Ⅳ row   for sample 4

The first column: speckle image; The second column: ground truth of  2D displacement;

The other column: the predicted 2D displacement field by training Mode A ,B,C

## 4.Discussions and conclusion

In this paper, we propose an unsupervised CNN-Based DIC for 2D Displacement Measurement. The work structure consists of an encoder-decoder architecture , which utilizes Multi-scale Feature Extraction Block, dual-path correlation block and hybrid attention module to extract effective information with different scale ,position and channel for representation. The composite loss function is adopted that incorporates both the Mean Squared Error (MSE) and Pearson correlation coefficient. The speckle image warp model is created to transform the target speckle image to the corresponding predicted reference speckle image by predicted 2D displacement. The predicted reference speckle image and the actual reference speckle image are compared to enable supervised learning of the network .After using a small amount of ground truth to conduct supervised pre-training on the network, our method can use unsupervised learning mode for network training, which only needs reference speckle and target speckle patterns without ground truth data. We conducted several experiments to demonstrate the validity and robustness of the proposed method. our proposed DICNet can reduce the MAE and RMSE errors of the predicted displacement by 58.44% and 57.14%, respectively, compared to Zhao et al's method . The MAE and RMSE errors between the predicted displacement by the unsupervised learning and ground-truth are only 0.0637 pixels and 0.828 pixels, respectively.The experimental results demonstrate that our method can achieve can achieve accuracy comparable to the supervised methods.The proposed method has some limitations. We need to conduct supervised pre-training on the network in order to ensure unsupervised training of the network to converge in the right direction. The proposed DIC method based on unsupervised convolutional neural network is only suitable for 2D displacement measurement. In future work, we will investigate how to extend this method to the application scenario of strain measurement, and also explore the possibility of fully unsupervised learning without the need for pre-training.

**Acknowledgment**: supported by the National Natural Science Foundation of China (Grant no:11672162).

**Disclosures**:The authors declare no conflicts of interest.

**Data availability:** The pytorch code UHRNet will be released at the URL: https://github.com/fead1

The datasets are from reference[18]

(https://pan.baidu.com/s/1KzC9g_GIkvMnGFumDYGyBA?pwd=fd5x).

## REFERENCES

[1]  1.Yang J, Bhattacharya K. Augmented Lagrangian digital image correlation[J]. Experimental Mechanics, 2019, 59(2): 187-205

2.Min H G, On H I, Kang D J, et al. Strain measurement during tensile testing using deep learningbased digital image correlation[J]. Measurement Science and Technology, 2019, 31(1): 015014

3.Boukhtache S, Abdelouahab K, Berry F, et al. When deep learning meets digital image correlation[J]. Optics and Lasers in Engineering, 2021, 136: 106308.

4.Boukhtache S, Abdelouahab K, Bahou A, et al. A lightweight convolutional neural network as an alternative to DIC to measure in-plane displacement fields[J]. Optics and Lasers in Engineering, 2023, 161: 107367.

5.Ma C, Ren Q, Zhao J. Optical-numerical method based on a convolutional neural network for fullfield subpixel displacement measurements[J]. Optics Express, 2021, 29(6): 9137-9156.

6.Ma X, Ren Q, Zhao D, et al. Convolutional neural network based displacement gradients estimation for a full-parameter initial value guess of digital image correlation[J]. Optics Continuum, 2022, 1(10): 2195-2211

7.Duan X, Xu H, Dong R, et al. Digital image correlation based on convolutional neural networks[J]. Optics and Lasers in Engineering, 2023, 160: 107234

8.Lan S H, Su Y, Gao Z R, et al. Deep learning for complex displacement field measurement[J]. Science China Technological Sciences, 2022: 1-18.

9.Wang Y, Zhao J. DIC-Net: Upgrade the performance of traditional DIC with Hermite dataset and convolution neural network[J]. Optics and Lasers in Engineering, 2023, 160: 107278

10.J. Tang, J. Wu, J. Zhang, et.al, "Single-Shot Diffraction Autofocusing: Distance Prediction via an Untrained Physics-Enhanced Network," in IEEE Photonics Journal, vol. 14, no. 1, pp. 1-6, Feb. 2022, Art no. 5207106, doi: 10.1109/JPHOT.2021.3138548

11.Fan S, Liu S, Zhang X, et al. Unsupervised deep learning for 3D reconstruction with dual-frequency fringe projection profilometry[J]. Optics Express, 2021, 29(20): 32547-32567

12.Junchao Zhang, Jianbo Shao, Jianlai Chen, et.al, "PFNet: an unsupervised deep network for polarization image fusion," Opt. Lett. 45, 1507-1510 (2020)

13.hao Zhang, Dangjun Zhao, Jianlai Chen, et.al"Unsupervised learning for hyperspectral recovery based on a single RGB image," Opt. Lett. 46, 3977-3980 (2021).

14.Liu KX, Wu JC, He ZH, Cao LC. 4K-DMDNet: diffraction model-driven network for 4K computer-generated holography. Opto-Electron Adv ,2023,6, 220135.

15.Fischer P , Dosovitskiy A , Ilg E , et al. FlowNet: Learning Optical Flow with Convolutional Networks[J]. IEEE, 2016.

16.Hu J, Shen L, Sun G. Squeeze-and-excitation networks[C]//Proceedings of the IEEE conference on computer vision and pattern recognition. 2018: 7132-7141.

17.https://pytorch.org/docs/stable/generated/torch.nn.functional.grid_sample.html

18.Xiao Hong, Li Chengnan, Feng Mingchi,et.al.，Large Deformation Measurement Method of Speckle Images Based on Deep Learning，Acta Optica Sinica，2023，43（14）：1412001